\documentclass[aps,prb,10pt, twocolumn]{revtex4}
\usepackage[dvipdfm]{graphicx,color}% Include figure files
\usepackage{dcolumn}% Align table columns on decimal point
\usepackage{bm}% bold math
\usepackage{amsmath,amsthm,amssymb,mathrsfs}
\newcommand{\lt}{\left}
\newcommand{\rt}{\right} %%%%%%% 

\newcommand{\Ev}{{\bm E}}

\newcommand{\sgn}{\rm {sgn}}

\newcommand{\jsv}{{\jv_{\rm s}}} 
\newcommand{\jv}{{\bm j}}

\newcommand{\kv}{{\bm k}}

\newcommand{\pv}{{\bm p}}

\newcommand{\efs}{\epsilon_{{\rm F}\sigma}}

\newcommand{\qv}{{\bm q}}

\newcommand{\sigmam}{\bm \sigma}

\newcommand{\rv}{{\bm r}} 
\newcommand{\mv}{{\bm m}} 
 
\newcommand{\la}{\langle} 
\newcommand{\ra}{\rangle}

\newcommand{\nv}{\bm{n}} 

\newcommand{\tauv}{\bm {\tau}}

\newcommand{\ve}{\varepsilon}
\newcommand{\vef}{\varepsilon_{\rm F}}

\newcommand{\w}{\omega}

\newcommand{\nus}{\nu_{\sigma}}

\newcommand{\bs}{{\bar \sigma}}

\newcommand{\up}{\uparrow}
\newcommand{\down}{\downarrow}

\begin{document}
\title{Spin torques due to diffusive spin current in magnetic texture}

\author{Kazuhiro Hosono$^1$, Junya Shibata$^2$, Hiroshi Kohno$^3$, and Yukio Nozaki $^{4, 5}$}%$^{1}$, Peter Entel$^{2}$}
\affiliation{
$^1$International Center for Materials Nanoarchitectonics (WPI-MANA), 
National Institute for Materials Science (NIMS), Namiki 1-1, Tsukuba 305-0044, Japan
\\
$^2$Faculty of Science and Engineering, Toyo University, 
Kawagoe, Saitama,350-8585, Japan
\\
$^3$Graduate School of Engineering Science, Osaka, University, 
Toyonaka, Osaka 560-8531, Japan 
\\
$^4$Department of Physics, Keio University, 
Hiyoshi, Yokohama 223-8522, Japan
\\
$^5$JST, CREST, Sanbancho, Chiyoda-ku, Tokyo, 102-0075, Japan
}
\date{\today}
\begin{abstract}
 We present a microscopic theory of spin torque due to diffusive spin currents 
induced by spin accumulation.
 The obtained expression is a natural extension of the existing one due to 
\lq local' spin currents associated with ordinary electric currents, 
and is the reciprocal of the spin motive force which induces charge accumulation    
as studied recently [J. Shibata and H. Kohno, Phys. Rev. B{\bf 84}, 184408 (2011)]. 
 The result is applied to a domain wall motion in a nonlocal spin injection system, 
and the torque and force due to diffusive spin current are evaluated. 
\end{abstract}

\maketitle
\section{introduction}
 The control of the magnetization by using flow of the spin of conduction electrons 
continues to be one of major research topics in the field of spintronics 
since the pioneering works of Slonczewski \cite{Slonczewski1996} 
and Berger \cite{Berger1996} on multilayer structures.
 Fundamental physics behind these phenomena is understood as due to spin torque, 
which arises from transfer of the spin angular momentum between conduction electron and magnetization. 
 The idea was first applied by Berger to the current-induced domain wall motion 
\cite{Berger1978, Berger1984, Berger1992}, where the electron spin can follow the 
local magnetization when traversing the domain wall. 
 Since the theoretical details of the current-induced spin torques were exposed by phenomenological \cite{{Zhang04},{TNMS05}} 
and microscopic \cite{Kohno2006} analyses, 
it is well-known that there are reactive torque (also known as spin-transfer torque) 
\cite{BJZ98,Ansermet04,Li04,Nakatani04} 
and dissipative torque (also called $\beta$ term) 
\cite{{Zhang04},{TNMS05},{TSBB06},{Kohno2007},{Duine07_PRB},{PT07}} 
in the presence of spin relaxation of conduction electrons. 
 They are expressed by the first and the second terms of
\begin{align}
\tauv_{\rm sd}^{(\rm L)} 
= \frac{\hbar}{2e} \lt\{ (\jv^{\rm (L)}_{\rm s} \!\cdot\! \nabla) \, \nv 
  + \beta \, \nv\times (\jv^{\rm (L)}_{\rm s} \!\cdot\! \nabla) \, \nv \rt\} , 
\label{eq:torque1}
\end{align}
respectively. 
 Here $\nv$ is a unit vector representing the direction of the localized spin (magnetization), 
$\jsv^{\rm (L)} = (\sigma_{\up} -\sigma_{\down}) \Ev$ is a spin-polarized current (spin current) 
driven by a local electric field $\Ev$, with 
$\sigma_{\up}$ ($\sigma_{\down}$) being spin-resolved conductivity for majority-spin (minority-spin) electrons, 
and $\beta$ is a constant originating from spin relaxation.

 Though it is well recognized that 
the spin torque is induced by the spin polarized current,  
the case of diffusive spin current has not been addressed in previous studies. 
 A diffusive spin current is induced from spin accumulation 
and is distinguished from spin-polarized current induced by local electric field.
 Experimentally, the local spin-polarized current can be separated off from 
the diffusive spin current by the nonlocal spin injection technique \cite{Jedema2001}.
 By using this method, magnetization switching by pure spin current 
and depinning of domain walls assisted by diffusive spin current 
were carried out \cite {Yang2008,Ilgaz2010}.
 The spin torque due to diffusive spin current has been discussed theoretically 
in ferromagnet/nonmagnet/ferromagnet/ nanopillar systems 
\cite{Berger2001, Fert2004} and 
the contribution of the diffusive spin current has been estimated quantitatively 
\cite{Kimura2006, Yang2006}. 
 However, a general expression of the spin torque induced by diffusive spin currents 
in magnetic texture, such as magnetic domain walls, has not been derived microscopically.

 As a reciprocal to the current-induced spin torque, 
a spin motive force is known to be generated by the dynamics of magnetization texture 
\cite{Berger86,Volovik87,Stern92,BM07,Duine08,Tserkovnyak08},
and actually has been detected experimentally \cite{YBKXNTE09}.
 In this context, two of the present authors studied microscopically 
the spin and charge transport induced by magnetization dynamics \cite{Shibata2011}. 
 They obtained a \lq local' charge current, 
${\bm j}^{(\rm L)} = \sigma_{\rm s}{\bm E}_{\rm s}$, driven by a spin motive force field 
\cite{Berger86,Volovik87,Stern92,BM07,Duine08,Tserkovnyak08} 
\begin{align} 
 E_{{\rm s},i}=
\frac{\hbar}{2e} \lt\{\nv\cdot \lt(\partial_i \nv \times {\dot \nv}\rt)+\beta\dot \nv \cdot \partial_i \nv \rt\} , 
\label{eq:2}
\end{align}
where 
the first and the second terms correspond to the respective terms in Eq.~(\ref{eq:torque1}). 
 Moreover, they obtained a diffusive charge current, 
${\bm j}^{(\rm D)}$, arising from charge imbalance induced by inhomogeneous ${\bm E}_{\rm s}$. 
 This diffusive current, in turn, implies the existence of spin torques 
induced by the diffusive spin current as the reciprocal, 
and it is expected that this torque takes the form 
\begin{align}
\tauv_{\rm sd}^{(\rm D)}
= \frac{\hbar}{2e} \lt\{ (\jv^{\rm (D)}_{\rm s} \!\cdot\! \nabla) \, \nv 
      + \beta \, \nv\times (\jv^{\rm (D)}_{\rm s} \!\cdot\! \nabla) \, \nv \rt\},
\label{eq:torque2}
\end{align}
where ${\bm j}^{(\rm D)}_{\rm s}$ is the diffusive spin current.

 The purpose of this paper is to clarify microscopically the effects of diffusive spin current 
on the spin torque. 
 It is shown that the general expression of the current-induced spin torque 
is given by the sum of Eq.~(\ref{eq:torque1}) and Eq.~(\ref{eq:torque2}). 
 The result is applied to a domain wall motion in a spin-valve system with nonlocal spin injector, 
and the force and torque acting on the domain wall due to diffusive spin current are evaluated.

 The paper is organized as follows. 
 After describing the model in Sec.~II,  
we calculate in Sec.~III the spin torque in response to an external electromagnetic field. 
 In Sec.~IV, we apply the obtained result to a domain wall motion 
in a nonlocal spin valve structure and evaluate the diffusive spin current 
and the torque and force acting on the domain wall. 
 Conclusion is given in Sec.~V.  
 Details of the calculation are given in Appendices.

%%%%%%%%%%%
%%%%%%%%%%%
\section{Model}
 We consider a ferromagnetic conductor described by the $s$-$d$ model 
which consists of conducting $s$-electrons and localized $d$-spins. 
 The Lagrangian for $s$-electrons is given by 
\begin{align}
 L_{\rm el}
&= \int d{\bm r}~ c^\dagger \lt(i\hbar \frac{\partial}{\partial t} + \frac{\hbar^2}{2m}\nabla^2 + \vef-V_{\rm imp}\rt) c 
          - H_{\rm sd}, 
\\
 H_{\rm sd} &= -M\int d{\bm r}~\nv ({\bm r}) \cdot (c^\dagger{\bm \sigma}c)_{x} , 
\end{align}
where $c^{\dagger}(x) = (c^{\dagger}_{\uparrow}(x),c^{\dagger}_{\downarrow}(x))$ 
is the electron creation operator at $x = (t,{\bm r})$,  
$\varepsilon_{\rm F} $ is the Fermi energy, 
$M$ is the $s$-$d$ exchange coupling constant, 
${\bm n}({\bm r})$ is a static but spatially-varying unit vector representing the direction of $d$-spin, 
which is treated as a classical spin, and 
${\bm \sigma}$ is a vector of Pauli matrices. 
 We assume that $s$-electrons are interacting with both non-magnetic and magnetic impurities, whose potential is modeled by  
\begin{align}
V_{\rm imp} &=u\sum_{i}\delta(\rv-{\bm R}_i)+u_{\rm s}\sum_{j}
{\bm S}_{j} \!\cdot \sigmam \, \delta(\rv-{\bm R}_{ j}'),
\label{eq:Vimp}
\end{align}
where $u$ ($u_{\rm s}$) is the strength of normal (magnetic) impurities, 
which introduce momentum relaxation (spin-relaxation) processes, 
and ${\bm R}_{i}$ (${\bm R}_{j}'$) is the position of normal (magnetic) impurities.

 To treat magnetization texture, 
we introduce a local spin frame for conduction electrons  
and take the $d$-spin direction $\nv(\rv)$ as the spin quantization axis. 
\cite{{KMP77},{Volovik87},{TF94}} 
 We define a new electron operator in the rotated frame by $a=Uc$, 
where $U$ is a 2 $\times$ 2 unitary matrix diagonalizing $H_{\rm sd}$, namely, 
it satisfies $U^\dagger (\nv \!\cdot\! \sigmam ) U=\sigma^z$. 
 It is convenient to choose $U = \mv \!\cdot\! \sigmam$ with 
\begin{align}
 \mv = \lt(\sin\frac{\theta}{2}\cos\phi,\sin\frac{\theta}{2}\sin\phi,\cos\frac{\theta}{2}\rt) , 
\end{align}
where $\theta$ and $\phi$ are ordinary polar coordinates parametrizing the direction of $\nv$.
The space and time derivative of the electron operator 
gives SU(2) gauge field 
\begin{align}
 A_\mu = -iU^\dagger \lt(\partial_\mu U\rt) = A^\alpha_\mu \sigma^\alpha, 
\end{align}
as $\partial_\mu c=U\lt(\partial_\mu +iA_\mu\rt)a$, 
which represents the spatio-temporal variation of magnetization. 
\cite{note1}
 The Lagrangian in the rotated frame is then given by 
$L=\tilde{L}_{\rm el}-H_{\rm e-A}$, 
\begin{align}
&\tilde{L}_{\rm el} = \int d\rv~ a^{\dagger}
\left[i\hbar
\frac{\partial}{\partial t}+
\frac{\hbar^2}{2m}\nabla^{2} + \varepsilon_{\rm F} - \tilde{V}_{\rm imp} + M \sigma^{z}
\right]a, \\
&H_{\rm e-A} 
= -\frac{\hbar}{e} \int d\rv~\tilde{j}^\alpha_\mu A^{\alpha}_\mu 
+\frac{\hbar^{2}}{2m} \int d\rv~A^{\alpha}_{i}A^{\alpha}_{i}
a^{\dagger}a ,
\end{align}
where 
$\tilde j^\alpha_\mu = (\tilde{\rho}^\alpha, \, \tilde {\bm j}^\alpha)$ 
is a four current density 
representing spin and spin-current densities 
(\lq\lq paramagnetic'' component) in the rotated frame, 
which are given by 
\begin{align}
 \tilde{\rho}^\alpha
&= -e a^{\dagger} \sigma^{\alpha} a  \ \ (\ = \, \tilde{j}^\alpha_0), 
\\
 \tilde{\bm j}^\alpha 
&= -e\frac{\hbar}{2mi} \left(a^{\dagger}\sigma^{\alpha}\nabla a 
 - (\nabla a^{\dagger}) \sigma^{\alpha}a\right) . 
\label{eq:js_tilde}
\end{align}
The spin part of the impurity potential 
$\tilde{V}_{\rm imp}$ is expressed as 
$S^{\alpha}_{j}(c^{\dagger}\sigma^{\alpha}c) 
=\tilde{S}^{\alpha}_{j}(t)(a^{\dagger}\sigma^{\alpha}a)$, 
where 
$\tilde{S}_{j}^{\alpha}(t) = {\cal R}^{\alpha\beta}({\bm R}_{j}',t)S^{\beta}_{j}$ 
is the impurity spin in the rotated frame 
with \cite{Kohno2007}
\begin{align}
{\cal R}^{\alpha\beta}= 2m^{\alpha}m^{\beta}-\delta^{\alpha\beta}
\end{align}
being a $3\times 3$ orthogonal matrix representing the same rotation as $U$. 
 We take a random average over impurity positions and for magnetic impurities 
we take a quenched average for the impurity spin direction 
as $\overline{\tilde{S}^\alpha_i}=0$ and 
\begin{align}
  \overline{\tilde S_i^\alpha \tilde S_j^\beta} 
= \delta_{ij} \delta^{\alpha\beta} \times \left\{ \begin{array}{cc} 
  \overline{S_\perp^2} & (\alpha, \beta = x,y) \\ 
  \overline{S_z^2}     & (\alpha, \beta = z) 
  \end{array} \right.  . 
\end{align}
 The anisotropy axis of the impurity spin is defined in reference to the rotated frame.

 In the following calculation, we use the impurity-averaged Green's function 
\begin{align}
 G_{\kv\sigma}(z)=\frac{1}{z-\ve_{\kv}+\ve_{{\rm F}\sigma}+i\gamma_\sigma {\sgn}({\rm Im} z)}
\label{eq:G}
\end{align}
where $\ve_{\kv}={\hbar^2\kv^2}/{2m}$, 
$\varepsilon_{{\rm F}\sigma}=\vef+\sigma M$. 
 The subscript $\sigma= \uparrow,\downarrow$ represents 
the majority and minority spins, respectively, 
and corresponds to $\sigma = +1,-1$ 
in the formula (and to $\bar{\sigma} = \downarrow, \uparrow$ or $-1,+1$). 
 The damping rate $\gamma_\sigma$ is evaluated 
in the first Born approximation as 
\begin{align}
\gamma_{\sigma} = \frac{\hbar}{2\tau_{\sigma}} 
= \pi ( \tilde{\Gamma}_{1}\nu_{\sigma} + \tilde{\Gamma}_{2}\nu_{\bar{\sigma}} ),
\label{eq:damping}
\end{align}
where $\nu_{\sigma} = mk_{{\rm F}\sigma}/2\pi^{2}\hbar^{2}$ 
is the density of states at $\varepsilon_{{\rm F}\sigma}$ 
with $k_{{\rm F}\sigma} = \sqrt{2m\varepsilon_{{\rm F}\sigma}}/\hbar$ and 
\begin{align}
\tilde{\Gamma}_{1} &= n_{\rm i}u^{2} + n_{\rm s}u^{2}_{\rm s}
\overline{S^{2}_{z}}, 
\\
\tilde{\Gamma}_{2} &= 2n_{\rm s}u^{2}_{\rm s}\overline{S^{2}_{\perp}}
\end{align}
with $n_{\rm i}$ ($n_{\rm s}$) being the concentration of 
normal (magnetic) impurities. 
The first and second terms in Eq.~(\ref{eq:damping}) come from 
spin-conserving and spin-flip scattering processes, respectively. 
In this paper, 
we assume weak impurity scattering, $\gamma_\sigma \ll \efs$.

%%%%%%%%%%%%%%%%%%%%%%%%%%%%%%%%%%%%
\section{Calculation of spin torque}

 Dynamics of the magnetization is described by the Landau-Lifshitz-Gilbert 
equation, \cite{Kohno2006}
\begin{align}
\dot\nv =\gamma_0 {\bm H}_{\rm eff}\times \nv +\alpha_0{\dot \nv\times \nv} +\tauv'_{\rm sd},
\end{align}
where $\gamma_0 {\bm H}_{\rm eff}$ and $\alpha_0$ are effective magnetic field and 
a Gilbert damping constant, respectively, in the absence of conduction electrons.
 Effects of conduction electrons are contained in the spin torque density, 
\begin{align}
\tauv'_{\rm sd}=
\frac{a^3}{\hbar S} M \nv\times \la \sigmam (\rv)\ra_{\rm ne} 
\end{align}
which comes from $H_{\rm sd}$. (Here, $a$ is the lattice constant and $S$ is the magnitude of $d$ spin.) 
 Our task is then to calculate $\la \sigmam (\rv)\ra_{\rm ne} = {\cal R} \la \tilde\sigmam (\rv)\ra_{\rm ne}$ 
in such nonequilibrium states with spin accumulation and the associated spin current 
in the presence of magnetization texture.

 To produce spin accumulation, we disturb the electron system by applying an electromagnetic 
scalar potential $\phi^{\rm em}$ and a vector potential ${\bm A}^{\rm em}$ which are 
time-dependent and inhomogeneous.
 The perturbation is described by the Hamiltonian 
\begin{align}
 H_{\rm em} 
= \int d\rv \left(
 \rho \, \phi^{\rm em} - {\bm j} \!\cdot\! {\bm A}^{\rm em} \right) 
= -\int d{\bm r} j_{\mu} A^{\rm em}_{\mu} . 
\end{align}
 Here we have introduced a four-component notation for the gauge potential,  
$A^{\rm em}_{\mu}=(-\phi^{\rm em},{\bm A}^{\rm em})$,  
and the charge/current density, $j_\mu = \tilde{j}_{\mu} - (e\hbar/m) 
 \delta_{\mu i} A^{\alpha}_{i} (a^{\dagger}\sigma^{\alpha}a)$, 
with
\begin{align}
& \tilde{j}_{0} = \tilde{\rho} = -ea^{\dagger}a , 
\\
& \tilde{\bm j} = -e\frac{\hbar}{2mi} ( a^{\dagger}\nabla a - (\nabla a^{\dagger}) a ) . 
\end{align}
 Working with Fourier components, we calculate $\langle \tilde{\bm \sigma}(\rv) \rangle_{\rm ne}$ 
as a linear response to $A^{\rm em}_{\mu}$, 
\begin{align}
 \lt< \tilde \sigma^\alpha(\qv) \rt>_{\rm ne}
= \sum_{\qv'} L^\alpha_{\mu}(\qv,\qv';\omega) A^{\rm em}_{\mu}(\qv',\omega), 
\label{eq:tilde-s}
\end{align}
where $\alpha = x,y$ takes transverse components 
({\it i.e.,} perpendicular to the local magnetization, which is $\hat z$ in the rotated frame). 
 The response function $L^\alpha_\mu (\qv,\qv';\omega)$ is obtained from 
\begin{widetext}
\begin{align}
 L^\alpha_\mu(\qv,\qv';i\omega_\lambda) 
&= \int_0^{1/T} d\tau \, 
    {\rm e}^{i\omega_\lambda \tau} \lt< {\rm T}_\tau \tilde \sigma^\alpha(\qv,\tau) j_\mu (-\qv') \rt>
\nonumber \\
&=\int_0^{1/T} d\tau \, {\rm e}^{i\omega_\lambda \tau} \lt\{ 
   \lt< {\rm T}_\tau \tilde \sigma^\alpha(\qv,\tau) \tilde j_\mu (-\qv')\rt> 
 -  \delta_{\mu i} \frac{e\hbar}{m}\sum_{\pv}A_i^\beta(\pv) \lt< {\rm T}_\tau \tilde \sigma^\alpha (\qv,\tau) \tilde \sigma^\beta(-\qv'-\pv)\rt>\rt\}.
\end{align}
\end{widetext}
by the analytic continuation, $i\omega_{\lambda} \to \hbar \omega + i0$,  
where $T$ is the temperature and $\omega_\lambda$ is a bosonic Matsubara frequency. 
 In this paper, we limit ourselves to absolute zero, $T=0$. 
 The average $\langle \cdots \rangle $ is taken in the equilibrium state determined by 
$\tilde{L}_{\rm el}$. 
 The Fourier components of the spin and (paramagnetic-like) current densities are given by 
\begin{align}
 \tilde{\sigma}^{\alpha}(\qv) &= \sum_{\kv}a^{\dagger}_{\kv_{-}}\sigma^{\alpha}a_{\kv_{+}}, 
\\
 \tilde{j}_{\mu}(\qv) 
&= -e \sum_{\kv}v_{\mu} a^{\dagger}_{\kv_{-}}a_{\kv_{+}} , 
\end{align}
with  
$\kv_{\pm} = \kv\pm\qv/2$, $v_{0} = 1$ and $v_i = \hbar k_i/m$. 
\begin{figure}[t]
\begin{flushleft}
%%%\resizebox{8.5cm}{!}{\includegraphics[angle=0]{fig1.eps}}
\includegraphics[scale=0.28]{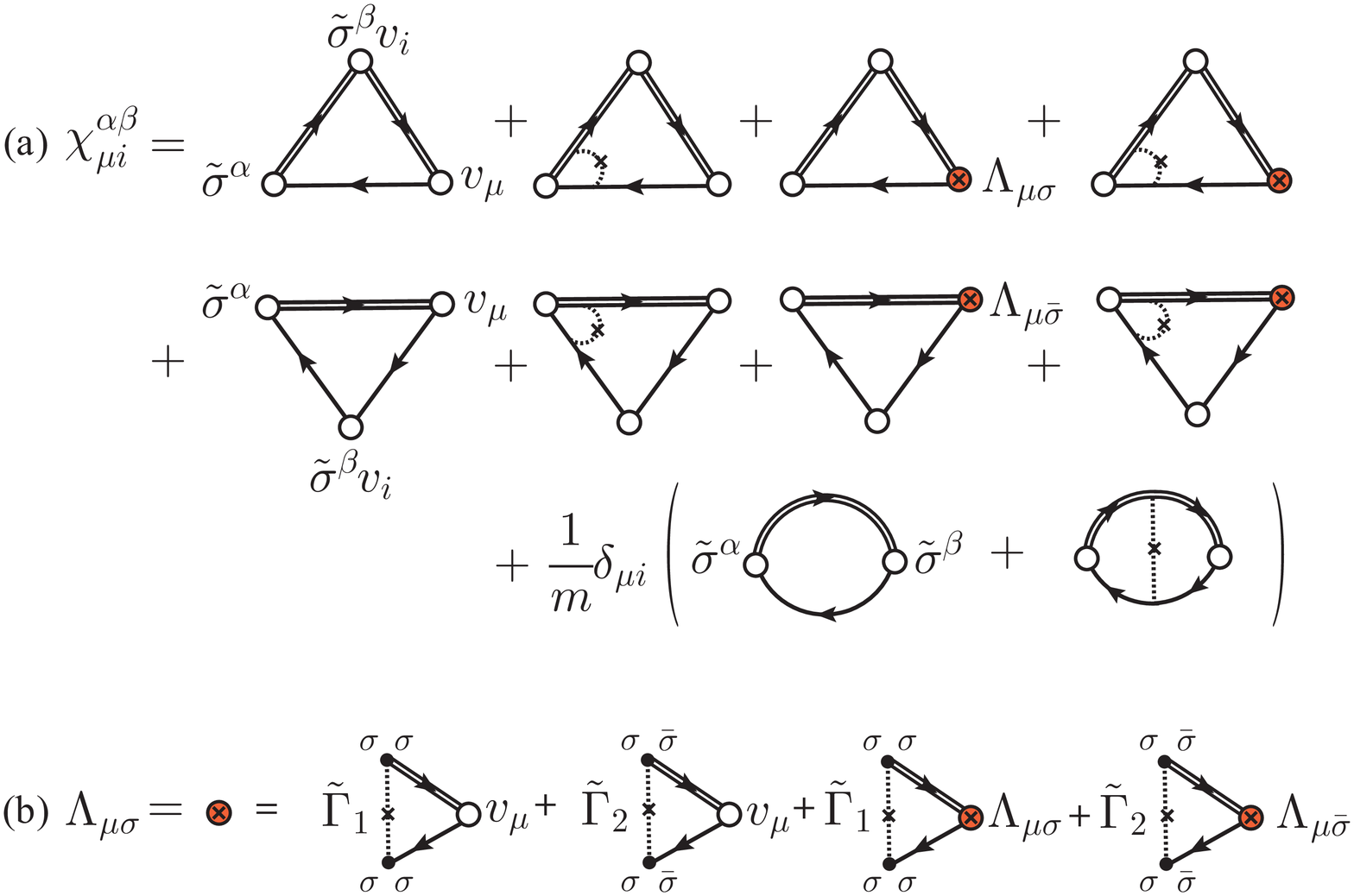}
\end{flushleft}
 \caption{(a) Diagrammatic representation of the coefficient $\chi^{\alpha\beta}_{\mu i}$ 
 of $A^{\rm em}_{\mu}$ and $A^\beta_{i}$ in the transverse spin polarization, $\la \tilde\sigma^\alpha (\rv)\ra_{\rm ne}$. 
 The solid (double solid) lines represent electron lines 
 carrying Matsubara frequency $i\ve_n$ ($i\ve_{n}+i\omega_{\lambda}$). 
 The dotted line represents scattering by impurities, 
 either non-magnetic or magnetic. 
 The filled vertex with a cross represents the ladder-type correction to the four-current vertex ($\Lambda_{\mu\sigma}$) 
as defined in (b).  
(b) Diagrammatic representation of Dyson equation for $\Lambda_{\mu\sigma}$. 
 The dotted line with $\tilde{\Gamma}_1$ ($\tilde{\Gamma}_2$) 
 denotes spin-nonflip (spin-flip) scattering.}
 \label{fig:1}
 \end{figure}
%%%%%%
 We extract an SU(2) gauge field $A^\beta_i$ from $L^\alpha_\mu$, and write as
\begin{align}
 L^{\alpha}_\mu(\qv,\qv';i\w_n) 
= e\hbar L^{\alpha \beta}_{\mu i}(\qv,\qv';i\w_n) A^\beta_{i}(\qv-\qv') . 
\label{eq:K-a-mu}
\end{align}
 The coefficient $L^{\alpha \beta}_{\mu i}$ is expressed by the diagrams shown 
in Fig.~\ref{fig:1} (a).

 Deferring the details of calculation to Appendices, we give the result here as 
\begin{align}
 L^{\alpha\beta}_{\mu i}(\qv,\qv';\w+i0) 
&= \frac{1}{M}
 ( \delta_{\perp}^{\alpha\beta} - \beta \varepsilon^{\alpha\beta})
 K^{\rm sc}_{i\mu} (\qv',\w+i0) , 
\label{eq:chi-ab-mui}
\end{align}
where 
$\delta_{\perp}^{\alpha\beta} = \delta_{\alpha\beta} - \delta_{\alpha z}\delta_{\beta z}$, 
$\varepsilon_{xy} = -\varepsilon_{yx}=1$ and 
\begin{align}
 \beta = \frac{\pi}{M} n_{\rm s}u_{\rm s}^{2}
(\overline{S^{2}_{\perp}}+\overline{S^{2}_{z}})(\nu_{\uparrow} + \nu_{\downarrow}) ,
\end{align}
is exactly the same as the coefficient of the $\beta$-terms of spin torque\cite{Kohno2006,Kohno2007} 
[Eq.~(\ref{eq:torque1})] and spin motive force\cite{Duine08,Shibata2011} [Eq.~(\ref{eq:2})]. 
 In Eq.~(\ref{eq:chi-ab-mui}), $K^{\rm sc}_{i\mu }$ is given by  
\begin{align}
&K^{\rm sc}_{i0}(\qv,\w+i0) = q_{i}\omega K^{\rm s} , 
\label{eq:Ki0}
\\
&K^{\rm sc}_{ij}(\qv,\w+i0)
=i\omega \left\{ \langle \sigma D\nu\rangle
\left(\delta_{ij}-\frac{q_{i}q_{j}}{q^2}
\right) -i\omega K^{\rm s}\frac{q_{i}q_{j}}{q^2} \right\} , 
\label{eq:Kij}
\end{align}
with 
\begin{align}
K^{\rm s} = \frac{\langle\sigma D\nu \bar{Y}\rangle+2\pi\tilde{\Gamma}_{2}
\langle \nu \rangle \langle \sigma D \nu \rangle}{Y_{\uparrow}Y_{\downarrow}+2\pi \tilde{\Gamma}_{2}
\langle Y \nu \rangle} . 
\label{eq:Ks}
\end{align}
 Here we have defined $Y_{\sigma} = D_{\sigma}q^{2} - i\omega$, 
$\la \sigma D\nu \ra = D_{\up}\nu_{\up}- D_{\down}\nu_{\down} $, 
$\la \sigma D\nu \bar{Y} \ra = D_{\up}\nu_{\up}Y_{\down}
- D_{\down}\nu_{\down}Y_{\up} $, 
$\la \nu \ra = \nu_{\up} + \nu_{\down}$, 
and $\la Y\nu \ra = Y_{\up}\nu_{\up} + Y_{\down}\nu_{\down}$.  
 Note that $K^{\rm sc}_{i\mu}$'s given by Eqs.~(\ref{eq:Ki0})-(\ref{eq:Ks}) agree with the result 
derived in Ref.~\onlinecite{Shibata2011} and describe the response of spin current to the electromagnetic field, 
\begin{align}
 j_{{\rm s},i}(\qv,\omega) 
&= e^2K^{\rm sc}_{i\mu}(\qv,\omega+i0) A^{\rm em}_{\mu}(\qv, \omega) . 
\label{eq:spin-current1}
\end{align}
 Here ${\bm j}_{\rm s}(\qv,\omega)$ is the Fourier component of the spin-current density, 
${\bm j}_{\rm s} \equiv \tilde {\bm j}^z$, the right-hand side being defined by Eq.~(\ref{eq:js_tilde}).

 Summarizing 
Eqs.~(\ref{eq:tilde-s}), (\ref{eq:K-a-mu}), (\ref{eq:chi-ab-mui}) and (\ref{eq:spin-current1}), 
we have 
\begin{align}
\lt<\tilde \sigma^\alpha (\qv)\rt>_{\rm ne}&=
\frac{e\hbar}{M}
 ( \delta_{\perp}^{\alpha\beta}-\beta\varepsilon^{\alpha\beta}) 
\nonumber\\
&\times\sum_{\qv'}A_i^{\beta}(\qv-\qv') 
K^{\rm sc}_{i\mu}(\qv',\w+i0) A^{\rm em}_\mu(\qv',\omega)\nonumber\\
&= \frac{\hbar}{eM}(\delta_{\perp}^{\alpha\beta}-\beta\varepsilon^{\alpha\beta})\sum_{\qv'}
{\bm A}^{\beta}(\qv-\qv')\cdot{\bm j}_{\rm s}(\qv',\omega).
\label{eq:sp}
\end{align}
 From the identities,\cite{Kohno2007} 
\begin{align}
 &{\cal R}^{\alpha\beta}\delta_{\perp}^{\beta\gamma} A^{\gamma}_{i} 
= -\frac{1}{2} \left(
\nv\times\partial_{i} \nv \right)^{\alpha}, \\
 &{\cal R}^{\alpha\beta} \varepsilon^{\beta\gamma} A^{\gamma}_{i} 
= -\frac{1}{2} \left(\partial_{i}\nv\right)^{\alpha}, 
\end{align}
we finally obtain the spin density as 
\begin{align}
\langle {\bm \sigma}({\bm r}) \rangle_{\rm ne} &= 
- \frac{\hbar}{2eM}
\left\{ \nv \times ({\bm j}_{\rm s} \!\cdot\! \nabla ) \, \nv - \beta \, ({\bm j}_{\rm s} \!\cdot\! \nabla ) \, \nv \right\} , 
\label{eq:spin}
\end{align}
and the spin-torque density as 
\begin{align}
 \tauv'_{\rm sd}(\rv,t) &= 
 \frac{a^3}{2eS} \left\{ ({\bm j}_{\rm s} \!\cdot\! \nabla) \, \nv
+ \beta \, \nv\times ({\bm j}_{\rm s} \!\cdot\! \nabla) \, \nv \right\} .
\label{eq:torque}
\end{align}
 The first and the second terms of $\tauv'_{\rm sd}(\rv,t)$ represent 
the spin-transfer torque and its dissipative correction ($\beta$ term), respectively,  
induced by ${\bm j}_{\rm s}$ given by Eq.~(\ref{eq:spin-current1}). 
 As shown in Ref. \onlinecite{Shibata2011}, ${\bm j}_{\rm s}$ is written as 
\begin{align}
 {\bm j}_{\rm s} &= {\bm j}_{\up}-{\bm j}_{\down} , 
\label{eq:js}
\\
 {\bm j}_{\sigma} &= 
 \sigma_{\sigma}{\bm E} - D_{\sigma} \nabla \rho_{\sigma} , 
\label{eq:j-sigma}
\end{align} 
where $\sigma_{\sigma}=e^2D_\sigma \nus$ is the electrical conductivity 
and $\rho_\sigma$ is the charge density of spin-$\sigma$ electrons.\cite{com_rho}
 Therefore, ${\bm j}_{\rm s}$ consists of two parts;
the ordinary (local) current, 
${\bm j}_{\rm s}^{\rm (L)} = (\sigma_\uparrow - \sigma_\downarrow){\bm E}$,  
and the diffusive (nonlocal) current, 
${\bm j}_{\rm s}^{\rm (D)} = -(D_\uparrow \nabla \rho_\uparrow - D_\downarrow \nabla \rho_\downarrow)$. 
 Equation (\ref{eq:torque}) is thus a natural extension of the previous current-induced torque 
which contained only ${\bm j}_{\rm s}^{\rm (L)}$. 
 It is reciprocal to the spin motive force 
induced by magnetization dynamics\cite{Shibata2011}. 
 These are the main results of this paper.

 It is noted that the spin current in Eq.~(\ref{eq:torque}) is within the two-current picture 
\cite{{Mott64},{Fert-Campbell68},{Son87},{Valet-Fert93}} 
and satisfies the spin \lq\lq continuity'' equation \cite{Shibata2011}
\begin{align}
\frac{\partial}{\partial t}\rho_\sigma +{\rm div}{\bm j}_{\sigma}=-\left(
\frac{\rho_\sigma}{\tau_{{\rm sf},\sigma}}-\frac{\rho_\bs}{\tau_{{\rm sf},\bs}}
\right), 
\label{eq:diffusion}
\end{align}
where $ \tau_{{\rm sf},\sigma}^{-1} = 2\pi \tilde{\Gamma}_{2}\nu_{\bs}/\hbar $
is the spin-flip rate for spin-$\sigma$ electrons. 
 Therefore, the diffusive part of the spin torque in Eq.~(\ref{eq:torque}) can be obtained 
by solving the ordinary diffusion equation in the two-current model 
obtained from Eqs.~(\ref{eq:j-sigma}) and (\ref{eq:diffusion}).

\section{Application}
 As an application, we consider a magnetic domain wall (DW) motion driven by a diffusive spin current 
 due to nonlocal spin injection in a lateral spin-valve system.  
 The system consists of two ferromagnetic wires, F1 and F2  
(with width $\omega_{\rm F}$ and thickness $d_{\rm F}$), and a non-magnetic 
wire, N (with width $\omega_{\rm N}$ and thickness $d_{\rm N}$). 
 F1 and F2 are separated by a distance $L$ and are connected by N, as shown in Fig. 2. 
%%%%%%%%%%%%
\begin{figure}[t]
%%%\resizebox{8cm}{!}{\includegraphics[scale=0.2]{fig2.eps}}
\includegraphics[scale=0.28]{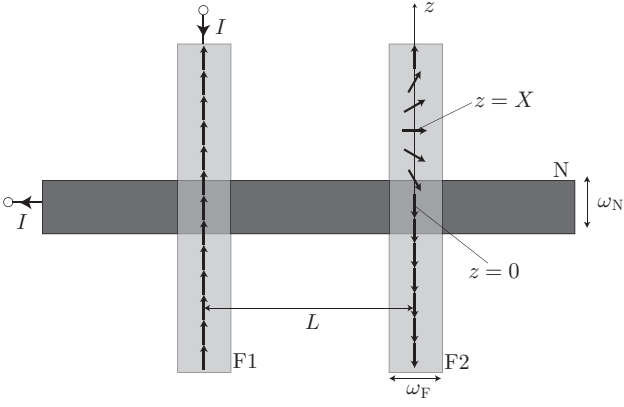}
\caption{Schematic illustration of a lateral spin-valve structure with two ferromagnetic leads (F1 and F2) 
connected by a non-magnetic lead (N). 
 The bias current $I$ flows from F1 to the left end of N, and produces a spin accumulation in N.  
 The induced diffusive spin current flows from the F1-N contact region 
to the N-F2 contact region and is injected into F2. 
 The injected spin current diffuses in F2 in positive and negative $z$-directions away from the contact. }
\label{fig:2}
\end{figure}
%%%%%%%%%%%%%%%%%%%% 
 There is a single DW in F2, while F1 is uniformly magnetized in $z$ direction. 
 When an electric current, $I$, is passed from F1 to the left end of N, as shown in Fig.~2, 
a spin-polarized current is injected from F1 into N and induces a non-equilibrium spin accumulation in N. 
 Then the spin current is injected into F2 and diffuses away from the injection point, 
in both positive and negative directions along the $z$-axis. 
% The magnitude of the spin accumulation decays as it 
%diffuses away from the injection point at $z=0$. 
 The spin-current density, $j_{\rm s}(z)$, flowing in F2 is then given by 
\begin{align}
 j_{\rm s}(z) =  j_{\up}(z)-j_{\down}(z)
&= -\sum_{\sigma}\sigma D_{\sigma} \partial_{z}\rho_{\sigma}\nonumber\\
&= \frac{\partial}{\partial z} \left(e\sum_{\sigma}\sigma D_{\sigma}
\nu_{\sigma}\delta\mu_{\sigma}\right), 
\end{align}
where $\delta \mu_{\sigma}(z)=-\rho_{\sigma}(z)/e\nu_{\sigma}$ 
is the nonequilibrium part of the chemical potential in F2. 
 It is obtained in 
 Refs. \onlinecite{Takahashi2003} and \onlinecite{Takahashi2008} as 
\begin{align}
\delta \mu_\sigma(z)={\bar \mu}+b^\sigma_2 e^{-{|z|}/{\lambda_{\rm F2}}} ,
\end{align}
where $\bar{\mu}$ and $b^{\sigma}_{2}$ are constants determined by 
the boundary condition, 
and $\lambda_{\rm F2}$ 
is the spin diffusion length in F2 given by \cite{Shibata2011}
\begin{align}
\frac{1}{\lambda^{2}_{\rm F2}}
=2\pi\tilde{\Gamma}_{2}
\frac{D_{\up}\nu_{\up}+D_{\down}\nu_{\down}}{D_{\up}D_{\down}}. 
\end{align}

 To evaluate the effect of spin torques on the DW motion, 
we assume a planar and rigid DW, a so-called one-dimensional model
\cite{Tatara2004}. 
 The DW motion is described by the relevant collective coordinates, 
the position $X$ of the DW center and the polarization angle $\phi_{0}$, 
which may depend on time. 
 Using these variables, the DW is expressed as $\nv(\rv,t) = \nv_{\rm w}(z-X(t))$, 
where 
\begin{align}
\nv_{\rm w}(z)&=\lt(\frac{\cos\phi_{0}}{\cosh(z/\lambda_{\rm w})}, 
\frac{\sin\phi_{0}}{\cosh (z/\lambda_{\rm w})}, 
\tanh\frac{z}{\lambda_{\rm w}}\rt)
\label{eq:ndw}
\end{align}
with $\lambda_{\rm w}$ being the DW width. 
 (We consider a tail-to-tail DW for simplicity.) 
 In the absence of the extrinsic pinning, 
the DW motion is described by the equation of motion 
\cite{{Tatara2004},{Tatara2008},{STK2011}}  
\begin{align}
\frac{\hbar SN_{\rm w}}{\lambda_{\rm w}}
\lt({\dot \phi}_{0}+\alpha\frac{\dot X}{\lambda_{\rm w}}\rt) &= F_{\rm el}, 
\\
\frac{\hbar SN_{\rm w} }{\lambda_{\rm w}}
\lt({\dot X}-\alpha \lambda_{\rm w}
{\dot \phi}_{0}\rt) &= \frac{K_{\perp}S^2N_{\rm w}}{2}\sin2\phi_{0}+ T_{{\rm el},z}, 
\end{align}
where $K_{\perp}$ is the hard-axis anisotropy constant, and 
$N_{\rm w}={2\lambda_{\rm w} A}/{a^3}$ 
is the number of spins in the wall with 
$A=\omega_{\rm F} d_{\rm F}$ being cross sectional area of F2 
and 
$a$ being the lattice constant. 
 The force $F_{\rm el}$ and torque $T_{{\rm el},z}$ are defined by 
\cite{Tatara2004, Tatara2008, STK2011}, 
\begin{align}
 F_{\rm el}&=-M\int d\rv~
 \partial_z\nv_{\rm w}(z-X(t))\cdot \la\sigmam(\rv)\ra_{\rm ne} , 
\label{eq:F}
\\ 
 T_{{\rm el},z}&=-M\int d\rv~
 \left[\nv_{\rm w}(z-X(t))\times \la\sigmam
 (\rv) \ra_{\rm ne}\right]_{z}. 
\label{eq:T}
\end{align}
 By substituting the spin polarization given by Eq.~(\ref{eq:spin}) into 
Eqs.~(\ref{eq:F}) and (\ref{eq:T}) and putting $\nv$ in $\nv_{\rm w}$, 
we obtain 
\begin{align}
&F_{\rm el}=-\beta \frac{\hbar}{2e}\frac{A}{\lambda_{\rm w}\lambda_{\rm F2}}
 C {\cal F}(\tilde{X};\xi), \\
&T_{{\rm el},z}
= -\frac{\hbar}{2e}\frac{A}{\lambda_{\rm F2}}
 C {\cal F}(\tilde{X};\xi), 
\label{eq:ft2}
\end{align}
where $C = e\sum_{\sigma}\sigma\nu_{\sigma}D_{\sigma}b_{2}^{\sigma}$ 
and 
\begin{align}
 {\cal F}(\tilde{X};\xi) 
= 2\xi\int_{-\infty}^{\infty}dz \, e^{-|z+\tilde{X}|/\xi} \frac{\sinh z}{\cosh^{3}z}, 
\end{align}
with $\tilde{X} = X/\lambda_{\rm w}$ and 
$\xi=\lambda_{\rm F2}/\lambda_{\rm w}$. 
 As pointed out in Ref. \onlinecite{Takahashi2008}, 
a large spin-current injection from {\rm N} into {\rm F2} is possible in a device 
with a tunnel junction at the injection part and a metallic contact with a strong spin absorber 
at the detection part of the nonlocal spin valve structure. 
 In such a case, we have 
\begin{align}
 C \simeq -\frac{\lambda_{{\rm F2}}}{A_{\rm J}}P_{1}I e^{-{L}/{\lambda_{\rm N}}}, 
\label{eq:c}
 \end{align}
where $A_{\rm J} = \omega_{\rm N}\omega_{\rm F}$ 
is the contact area of the junction and $P_1$ is the spin polarization of F1 
and 
$\lambda_{\rm N}$ is the spin diffusion length in N. 
 Thus we obtain 
\begin{figure}[t]
%%%\resizebox{8cm}{!}{\includegraphics[angle=0]{fig3.eps}}
\includegraphics[scale=0.27]{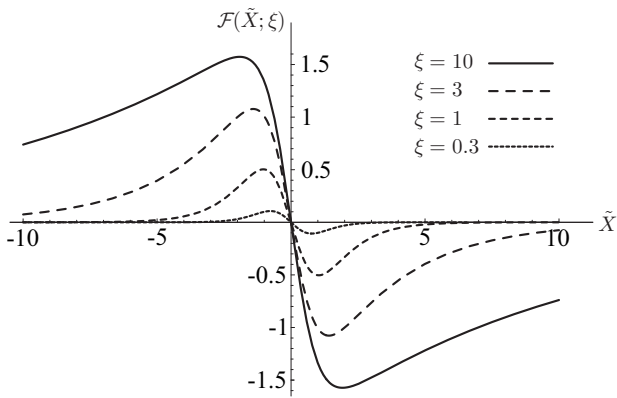}
\caption{Graph of ${\cal F}(\tilde{X};\xi)$ as a function of $\tilde{X} = X/\lambda_{\rm w}$ 
for several choices of $\xi=\lambda_{{\rm F2}}/\lambda_{\rm w}$. }
\label{fig:3}
\end{figure}
\begin{align}
&F_{\rm el} = \frac{\hbar}{2e}\frac{1}{\lambda_{\rm w}}\beta I^{(\rm D)}_{\rm s},
\\
&T_{{\rm el},z} = \frac{\hbar}{2e}I^{(\rm D)}_{\rm s}, 
\end{align}
where 
\begin{align}
 I^{(\rm D)}_{\rm s} = \frac{A}{A_{\rm J}}
 P_1 I e^{-{L}/{\lambda_{N}}} {\cal F}(\tilde{X};\xi) , 
\end{align}
is the effective spin current due to spin diffusion. 
 Figure \ref{fig:3} shows 
${\cal F}(\tilde{X};\xi)$ 
as a function of $\tilde{X}$ for several values of $\xi$. 
 Two (positive and negative) peaks are seen at around $|\tilde{X}| \simeq 1$ for each $\xi$, 
which corresponds to the distance between the wall center and the injection point. 
 This is due to properties of that diffusive spin current flows 
in gradient direction of the spin accumulation.  
 In a one-dimensional wire, 
the diffusive spin current flows in positive and negative $z$-directions 
and drives the wall in mutually opposite direction separating the injection point. 
 Note that its direction can be reversed by reversing the current direction 
without reversing the magnetization of F1, 
or by reversing the magnetization of F1 without reversing the current direction.  
 We also see that the force and torque acting on the DW  
strongly depends on the values of $\xi$. 
 For smaller $\xi$, the peak value of ${\cal F}$ is also smaller, 
because of the shorter spin-diffusion length. 
 These facts show that an efficient angular momentum transfer 
can be realized if we could generate the spin accumulation 
at the around edge of the DW and we use materials with large value of $\xi$,  
where the spin diffusion length $\lambda_{\rm F2}$ 
is much longer than the wall width.

%%%%%%%%%%%%%%%%%%%%%%%%%%%%%%%%%%%
\section{Conclusion}
 We have presented a microscopic theory of current-induced spin torque 
with emphasis on the role of diffusive spin current.   
% Effects of spin relaxation of electrons are taken into account. 
 The obtained torque consists of reactive and dissipative terms,     
and provides a natural extension of the well-established one 
due to \lq local' spin current which accompanies ordinary (Ohmic) electric current. 
 We applied the result to a domain wall in a lateral spin-valve system with nonlocal spin injection.  
 The effects on the domain wall (force and torque) depend strongly on the ratio 
of the spin diffusion length to the wall width, 
and increase when the spin diffusion length is much longer than the wall width.

 An interesting extension of the present theory would be to the study 
of modulation of the magnetization damping due to spin accumulation, 
as observed experimentally.\cite{Ando2008}  
 This will be reported elsewhere.

\section*{Acknowledgement}
 The authors are grateful to G. Tatara, A. Yamaguchi and K. Sekiguchi for valuable discussions.
This work is supported by Grant-in-Aid for Young Scientists B (No. 24710153) 
from Japan Society for the Promotion of Science and a JST, CREST. 
%%%%%%%%%%%%%%%%%%
%%%%%%%%%%%%%%%%%%
%%%%%%%%%%%%%%%%%%
%%%%%%%%%%%%%%%%%%
\appendix
\begin{widetext}

\section{Calculation of $L^{\alpha\beta}_{\mu i}$}

 In this Appendix, we evaluate the response function $L^{\alpha\beta}_{\mu i}$ in Eq.~(\ref{eq:K-a-mu}).   
 It is expressed as (see Fig.~1) 
\begin{align}
 L^{\alpha \beta}_{\mu i}(\qv,\qv';i\w_\lambda)
&= T\sum_n \sum_{\bm k} \biggl\{ 
           {\rm tr} [\, \Sigma^\alpha G^+_{{\bm k}+{\bm q}-{\bm q}'/2} \, v_i^+ \sigma^\beta 
                        G^+_{{\bm k}+{\bm q}'/2} \, \tilde \Lambda_\mu G_{{\bm k}-{\bm q}'/2} \, ] 
\nonumber \\
& \hskip 2cm
         + {\rm tr} [\, \Sigma^\alpha G^+_{{\bm k}+{\bm q}'/2} \, \tilde \Lambda_\mu G_{{\bm k}-{\bm q}'/2} 
                     \, v_i^- \sigma^\beta G_{{\bm k}-{\bm q}+{\bm q}'/2}  \, ] 
         + \frac{\delta_{\mu i}}{m} \, 
           {\rm tr} [\, \Sigma^\alpha G^+_{{\bm k}+{\bm q}/2} \, \sigma^\beta G_{{\bm k}-{\bm q}/2} \,] 
   \biggr\} ,
\label{eq:L_qq'}
\end{align}
where  $v^{\pm}_i=(\hbar/m)\lt( k_i \pm q_i /{2}\rt)$, 
and we have adopted a $2\times 2$ matrix notation, $(G)_{\sigma \sigma'} = G_\sigma \delta_{\sigma \sigma'}$. 
 We have included a ladder correction to the four-current vertex $v_\mu$ 
via $\tilde \Lambda_\mu \equiv v_\mu + \Lambda_\mu$ ($\Lambda_\mu$ being defined in Fig.~1(b)), 
and a first-order correction to the transverse-spin vertex, $\sigma^\alpha$ ($\alpha = x,y$), via  
\begin{align}
 \Sigma^\alpha &= \sigma^\alpha + \tilde{\Gamma}_{0}\sum_{\kv } G_{\kv-\qv/2} \, \sigma^\alpha \, G^+_{\kv+\qv/2} , 
\end{align} 
where $\tilde{\Gamma}_0 = n_{\rm i} u_0^2 - n_{\rm s} u_{\rm s}^2 \overline{S^{2}_{z}}$. 
 In the following calculation, we consider the adiabatic limit and put ${\bm q} = {\bm 0}$ in the Green's functions.  
 Namely, we consider 
$L^{\alpha \beta}_{\mu i}(\qv',i\w_\lambda) \equiv L^{\alpha \beta}_{\mu i}({\bm 0},\qv';i\w_\lambda)$, 
or using ${\bm q}$ for ${\bm q}'$, 
\begin{align}
 L^{\alpha \beta}_{\mu i}(\qv,i\w_\lambda)
&= T\sum_n \sum_{\bm k} \biggl\{ 
           {\rm tr} [\, \Sigma^\alpha G^+_{{\bm k}-} \, v_i \sigma^\beta 
                        G^+_{{\bm k}+} \, \tilde \Lambda_\mu G_{{\bm k}-} \, ] 
         + {\rm tr} [\, \Sigma^\alpha G^+_{{\bm k}+} \, \tilde \Lambda_\mu G_{{\bm k}-} 
                     \, v_i \sigma^\beta G_{{\bm k}+}  \, ] 
         + \frac{\delta_{\mu i}}{m} \, 
           {\rm tr} [\, \Sigma^\alpha G^+_{{\bm k}+} \, \sigma^\beta G_{{\bm k}-} \,] 
   \biggr\} ,
\label{eq:L_q}
\end{align}
where ${\bm k} \pm \equiv {\bm k} \pm {\bm q}/2$. 
 Note that, here and hereafter, ${\bm q}$ actually means ${\bm q}'$ of $A_\mu^{\rm em}({\bm q}')$. 
 We retain terms up to ${\cal O}(\omega, q^2)$, and to the next leading order 
with respect to the electron damping $\gamma_\sigma$ to see the spin-relaxation effects. 
 Equation (\ref{eq:L_q}) is written as 
\begin{align}
& L^{\alpha \beta}_{\mu i}(\qv,i\w_\lambda) 
= T\sum_{n} \sum_\sigma \lt(\delta_{\perp}^{\alpha \beta} + i\sigma \varepsilon^{\alpha \beta}\rt)
  \varphi_{\mu i, \sigma} (\qv; i\ve_n+i\w_\lambda,i\ve_n) , 
\end{align}
with 
\begin{align}
 \varphi_{\mu i, \sigma} (\qv; i\ve_n+i\w_\lambda,i\ve_n) 
&= (1+\Gamma_{\sigma})  \sum_{\kv}\biggl\{ 
 v^+_i  \tilde \Lambda_{\mu\sigma} G^{+}_{\kv-,\bs} G^{+}_{\kv+,\sigma} G_{\kv-,\sigma}  
 + v_{i}^{-}  \tilde \Lambda_{\mu\bs}  G^{+}_{\kv+,\bs} G_{\kv-,\bs} G_{\kv+,\sigma}  
 +  \frac{\delta_{\mu i}}{m} G^{+}_{\kv+,\bs} G_{\kv-,\sigma} \biggr\}  , 
\end{align}
where 
\begin{align}
 \Gamma_{\sigma} = \tilde{\Gamma}_0 \sum_{\kv } G^+_{\kv \bs} G_{\kv\sigma} . 
\end{align} 
 Performing the analytic continuation, 
$i\omega_{\lambda} \to \hbar\omega + i0$ 
and retaining terms up to the first order in $\omega$, we obtain 
\begin{align}
& L^{\alpha\beta}_{\mu i}(\qv, \omega + i 0) - L^{\alpha\beta}_{\mu i}({\bm q},0)
\nonumber\\
&=  \frac{\omega}{2\pi i}\sum_{\sigma}
 (\delta_{\perp}^{\alpha\beta}+i\sigma \varepsilon^{\alpha\beta})
 \left\{  \varphi^{(2)}_{\mu i, \sigma}(\qv;\omega,0)
-{\rm Re} \, \varphi^{(1)}_{\mu i, \sigma}(\qv;0,0) 
 - 2i\int_{-\infty}^{\infty} d\varepsilon f(\varepsilon) \left. 
  \frac{\partial}{\partial \omega}{\rm Im}\varphi^{(1)}_{\mu i, \sigma} (\qv;\ve_{+},\ve_{-})
  \right|_{\omega =0} 
 \right\} , 
\label{eq:w_linear}
\end{align}
where $\ve_{\pm} = \ve \pm \omega /2$ and $f(\varepsilon)$ is the Fermi distribution function.  
 The superscripts (1) and (2) indicate the analytic continuations, 
$G(i\ve_{n}+i\omega_{\lambda})G(i\ve_{n}) \to G^{\rm R} G^{\rm R}$ and $G^{\rm R} G^{\rm A}$, respectively. 
 The $\varphi^{(1)}_{\mu i, \sigma}$ and $\varphi^{(2)}_{\mu i, \sigma}$ are given by
\begin{align}
 \varphi^{(1)}_{\mu i, \sigma}(\qv;0,0) 
&= \lt( 1+\Gamma^{(1)}_{\sigma} \rt) 
 \lt\{ B_{\mu i, \sigma}(\qv) + B_{\mu i,\bs}(-\qv) 
 + \frac{\delta_{\mu i}}{m\tilde{\Gamma}_0} \, \Gamma^{(1)}_{\sigma} \rt\}, 
\\
 \varphi^{(2)}_{\mu i, \sigma}(\qv;\omega,0) 
&= \lt( 1+\Gamma^{(2)}_{\sigma} \rt)
 \lt\{ C_{\mu i,\sigma}(\qv)
 + C^{\ast}_{\mu i, \bs}(-\qv)
 + \Lambda^{(2)}_{\mu\sigma} ({\bm q},\omega) C_{0i,\sigma}(\qv) 
 + \Lambda^{(2)}_{\mu\bs} ({\bm q},\omega) C^\ast_{0i,\bs}(-\qv) 
 + \frac{\delta_{\mu i}}{m\tilde{\Gamma}_0} \, \Gamma^{(2)}_{\sigma} \rt\} , 
\label{eq:phi_2}
\end{align}
where 
\begin{align}
 B_{\mu \nu, \sigma}(\qv) 
&= \sum_{\kv} v_\mu v_\nu 
 G^{\rm R}_{\kv-,\bs} G^{\rm R}_{\kv+,\sigma} G^{\rm R}_{\kv-,\sigma} ,
\label{eq:B}
\\
 C_{\mu\nu,\sigma}(\qv)
&= \sum_{\kv} v_\mu v_\nu 
 G^{\rm R}_{\kv-,\bs} G^{\rm R}_{\kv+,\sigma} G^{\rm A}_{\kv-,\sigma} .
\label{eq:C}
\end{align}
 Since the terms of our interest are ${\cal O}(\gamma_\sigma^0)$ or lower order in $\gamma_\sigma$,  
the four-current vertex correction $\Lambda_\mu$ is relevant only in $\varphi^{(2)}$ 
and has been dropped in $\varphi^{(1)}$. 
 The $\Lambda^{(2)}_{\mu\sigma} ({\bm q},\omega)$ is evaluated in the diffusion approximation 
by retaining ${\bm q}$ and $\omega$ in a usual way, with a result \cite{Shibata2011}  
\begin{align}
&\Lambda_{0\sigma}^{(2)} =  
 \frac{Y_{\bs} + 2\pi \tilde{\Gamma}_{2} \langle\nu\rangle}
 {Y_{\up}Y_{\down}+2\pi\tilde{\Gamma}_{2}\langle Y\nu \rangle}
 \!\cdot\! \frac{1}{\tau_\sigma} , 
\ \ \ \ \ \ \  
\Lambda^{(2)}_{i\sigma} 
= -iq_{i} \frac{D_{\sigma}Y_{\bs}+2\pi\tilde{\Gamma}_{2}\langle D\nu \rangle}
 {Y_{\up}Y_{\down}+2\pi\tilde{\Gamma}_{2}\langle Y\nu \rangle} \!\cdot\! \frac{1}{\tau_\sigma} , 
\end{align}
where
$Y_{\sigma}=D_{\sigma}q^{2}-i\omega$, 
$\langle \nu \rangle = \nu_{\up}+\nu_{\down}$, 
$\langle Y\nu \rangle = Y_{\up}\nu_{\up} + Y_{\down}\nu_{\down}$ 
and 
$\langle D\nu \rangle = D_{\up}\nu_{\up} + D_{\down}\nu_{\down}$, 
with $D_{\sigma} = v^{2}_{{\rm F}\sigma}\tau_{\sigma}/3$ ($v_{{\rm F}\sigma} = \hbar k_{{\rm F}\sigma}/m$) 
being a diffusion constant.

 In $B_{\mu \nu, \sigma}$ and $C_{\mu \nu, \sigma}$, 
${\bm q}$ is retained up to the first order whereas $\omega$ is set to zero, $\omega =0$. 
 These are calculated in Appendix B, and the results lead to 
 ${\rm Re} \, \varphi^{(1)}_{\mu i, \sigma} = {\cal O}(\gamma_\sigma)$ and 
\begin{align}
 \varphi^{(2)}_{\mu i, \sigma}(\qv;\omega,0) 
&= -\frac{1}{2M}(1+i\sigma\beta)\sigma
 \left(
 \Pi^{(2)}_{i\mu,\sigma}+\Pi^{(2)}_{i0,\sigma}
 \Lambda^{(2)}_{\mu \sigma}
 \right)+\frac{1}{2M}(1+i\sigma\beta)\sigma
 \left(
 \Pi^{(2)}_{i\mu ,\bs}+\Pi^{(2)}_{i0,\bs}
 \Lambda^{(2)}_{\mu \bs}
 \right) 
\nonumber\\
&{} \ \ \ \  
 + \delta_{\mu i} \frac{i\pi \sigma}{2mM^2} (\rho_{\rm s} - M \nu_+ )  
% + \delta_{\mu 0} \frac{i\pi q_i}{4mM^3} (\rho_{\rm s} - M \nu_+ )  , 
\label{eq:phi2}
\end{align}
where 
\begin{align}
 \Pi_{\mu\nu,\sigma}^{(2)} = \sum_{\kv} v_{\mu} v_\nu G^{\rm R}_{\kv + ,\sigma} G^{\rm A}_{\kv-,\sigma} . 
\label{eq:Pi}
\end{align} 
 The last term in the braces of Eq.~(\ref{eq:w_linear}) is calculated as 
\begin{align}
 -2i\int^{\infty}_{-\infty}d\ve f(\ve )
 \left. \frac{\partial}{\partial \omega}{\rm Im} \varphi^{(1)}_{\mu i, \sigma} \right|_{\omega=0}
&\simeq
 -2 i \frac{\delta_{\mu i}}{m} \int^0_{-\infty} d\ve \, {\rm Im} 
 \sum_{\kv} \left[ 
(G^{\rm R}_{\kv\bs})^{2}G^{\rm R}_{\kv\sigma} - G^{\rm R}_{\kv\bs}(G^{\rm R}_{\kv\sigma})^{2}
\right] 
\nonumber\\
&= -\delta_{\mu i} \frac{i \pi \sigma}{2mM^2} (\rho_{\rm s} - M \nu_+ )  , 
\end{align}
which cancels the terms in the second line of Eq.~(\ref{eq:phi2}). 
 Therefore, we obtain 
\begin{align}
 L^{\alpha\beta}_{\mu i }(\qv,\w+i0) - L^{\alpha\beta}_{\mu i}({\bm q},0)
&= \frac{1}{M} (\delta_{\perp}^{\alpha\beta}-\beta\varepsilon^{\alpha\beta})
\frac{i\w}{2\pi}\sum_{\sigma}\sigma
\lt( \Pi^{(2)}_{i\mu,\sigma} + \Pi^{(2)}_{i0,\sigma}\Lambda^{(2)}_{\mu\sigma} \rt)
\nonumber \\
&= \frac{1}{M} 
 (\delta_{\perp}^{\alpha\beta}-\beta\varepsilon^{\alpha\beta})
 K^{\rm sc}_{i\mu}(\qv,\w+i0) , 
\end{align}
or, in the original notation in Eq.~(\ref{eq:L_qq'}), 
\begin{align}
 L^{\alpha \beta}_{\mu i}(\qv,\qv';\w+i0)
&\simeq  L^{\alpha \beta}_{\mu i}({\bm 0},\qv';0) 
 + \frac{1}{M} (\delta_{\perp}^{\alpha\beta}-\beta\varepsilon^{\alpha\beta})
   K^{\rm sc}_{i\mu}(\qv',\w+i0) , 
\label{eq:L}
\end{align}
\end{widetext}
where 
\begin{align}
 K^{\rm sc}_{i\mu}(\qv,\w+i0) 
&= \frac{i\w}{2\pi}\sum_{\sigma}\sigma 
 \lt( \Pi^{(2)}_{i\mu,\sigma} + \Pi^{(2)}_{i0,\sigma} \Lambda^{(2)}_{\mu \sigma} \rt)
\end{align}
is nothing but the electromagnetic response function of spin current $j_{{\rm s},i}$ 
(see Eq.~(\ref{eq:spin-current1})) obtained in Ref.~\onlinecite{Shibata2011}. 
 The first term of Eq.~(\ref{eq:L}) gives a spin torque 
due to diamagnetic spin current, which will be reported elsewhere.

\section{The $\kv$-integrals}

 The ${\bm k}$ integrals of the product of Green's functions are given here. 
 We retain low-order contributions with respect to $\gamma_\sigma$ and ${\bm q}$.

 Equation (\ref{eq:Pi}) is calculated as \cite{Shibata2011} 
\begin{align}
&\Pi^{(2)}_{00,\sigma} \simeq 2\pi \nu_{\sigma}\tau_{\sigma} , 
\\
&\Pi^{(2)}_{i0,\sigma} \simeq -2\pi iq_{i}D_{\sigma}\nu_{\sigma}\tau_{\sigma} , 
\\
&\Pi^{(2)}_{ij,\sigma} \simeq 2\pi D_{\sigma}\nu_{\sigma}\delta_{ij} , 
\end{align}
where 
$D_{\sigma} = v^{2}_{{\rm F}\sigma}\tau_{\sigma}/3$ with  
$v_{{\rm F}\sigma} = \hbar k_{{\rm F}\sigma}/m$. 
 
 The $C_{ij,\sigma}(\qv)$ in Eq.~(\ref{eq:C}) is calculated as 
\begin{align}
 C_{ij,\sigma}(\qv) 
&\simeq  C_{ij,\sigma}({\bm 0}) 
= \sum_{\kv} v_i v_j G^{\rm R}_{\kv \bs} G^{\rm R}_{\kv \sigma} G^{\rm A}_{\kv \sigma} 
\nonumber \\
&= \frac{1}{2(\sigma M - i\gamma_+)} 
  \sum_{\kv} v_iv_j G^{\rm R}_{\kv \sigma} (G^{\rm R}_{\kv \bs} - G^{\rm A}_{\kv \sigma}) 
\nonumber\\
&\simeq  -\frac{\sigma}{2M}\left( 1 + i \sigma \frac{\gamma_{+}}{M} \right) \Pi^{(2)}_{ij,\sigma} 
 + \delta_{ij}\frac{i\pi}{4mM^2} \sigma\rho_{\rm s} ,
\end{align} 
where $2\gamma_+ = \gamma_{\up} + \gamma_{\down}$, and 
$\rho_{\rm s} = n_{\up}-n_{\down}$ is the spin density with 
$n_{\rm \sigma} = 2\varepsilon_{{\rm F}\sigma}\nu_{\sigma}/3$. 
 In the last line, we have used Eq.~(\ref{eq:Pi}).

 The $C_{i0,\sigma}({\bm q})$ in Eq.~(\ref{eq:B}) is calculated as 
\begin{align}
 C_{i0,\sigma} 
&= \sum_{\kv}v_i G^{\rm R}_{\kv-,\bs}  G^{\rm R}_{\kv+,\sigma}  G^{\rm A}_{\kv-,\sigma} 
\nonumber\\
&= \frac{1}{2(\sigma M - i\gamma_+)} 
  \sum_{\kv}v_i  G^{\rm R}_{\kv+,\sigma} (G^{\rm R}_{\kv-,\bs} - G^{\rm A}_{\kv-,\sigma}) 
\nonumber\\
&\simeq  -\frac{\sigma}{2M} \left(1+i\sigma \frac{\gamma_{+}}{M} \right) \Pi^{(2)}_{i0,\sigma} . 
\end{align}

\begin{widetext}
 Finally, Eq.~(\ref{eq:phi_2}) is calculated, 
with $\Gamma^{(2)}_{\sigma} = -i\pi\sigma \tilde{\Gamma}_0 \nu_+/2M + {\cal O}(\gamma)$, as   
\begin{align}
& (1+ \Gamma^{(2)}_{\sigma})(C_{\mu i,\sigma} + \Lambda^{(2)}_{\mu \sigma} C_{0i,\sigma}) 
\nonumber \\
&=
-\frac{\sigma}{2M}
 \left( 1 - i \sigma \frac{\pi}{2M} \tilde{\Gamma}_{0} \nu_{+}\right)
 \left( 1 + i \sigma \frac{\gamma_{+}}{M}  \right)
 \left( \Pi^{(2)}_{i\mu,\sigma} + \Pi^{(2)}_{i0,\sigma} \Lambda^{(2)}_{\mu\sigma} \right)
 + \delta_{\mu i} \frac{i\pi \sigma}{4mM^2} \rho_{\rm s} 
\nonumber\\
&\simeq 
 -\frac{\sigma}{2M}\left(1+i\sigma\beta\right)
   \left(\Pi^{(2)}_{i\mu,\sigma} + \Pi^{(2)}_{i0,\sigma} \Lambda^{(2)}_{\mu\sigma}\right)
 + \delta_{\mu i} \frac{i\pi \sigma}{4mM^2} \rho_{\rm s} 
\label{eq:Bimu}
\end{align}
where 
\begin{align}
\beta &= \frac{\gamma_+}{M} - \frac{\pi}{2M} \tilde{\Gamma}_0 \nu_+ 
= \frac{\pi}{M}n_{\rm s}u_{\rm s}^{2}
   \left( \overline{S^{2}_{\perp}} + \overline{S^{2}_{z}} \right) (\nu_{\up}+\nu_{\down}). 
\end{align}
 This leads to Eq.~(\ref{eq:phi2}). 

\end{widetext}

%%%%%%%%%%%%%%%%%%%%%%%%%%%%%%%%%%%%%%

%\end{widetext}
%\bibliography{library}

\begin{thebibliography}{99}
\bibitem{note1}
We use Greek superscripts $\alpha, \beta$, ...for spin components, 
and Greek subscripts	 $\mu$ for components of four vectors, whose 
space (time) components are expressed by Latin subscripts $i, j$,...(by 0). 
Repeated indices generally imply summation.
\end{thebibliography}

\end{document}